\documentclass[nofootinbib,letterpaper,aps,amsmath,amsfonts,12pt]{revtex4}
\usepackage[matrix,line,graph,curve]{xy}
\begin{document}

\setlength\arraycolsep{2pt} 
\def\Pcm#1{{\mathcal{#1}}}
\def\nn{\nonumber}
\def\er#1{eqn.\eqref{#1}}
\def\fgref#1{fig.\ref{#1}}
\newcommand{\del}{\partial}
\newcommand{\re}{\Re \textup{e} \ }
\newcommand{\im}{\Im \textup{m} \ }
\newcommand{\td}{\textup{d}}
\newcommand{\tr}{\textup{tr}}
\def\busch#1{\check{#1}}
\def\bucsh#1{\check{#1}}
 
\title{Explicitly connecting T~and Buscher dualities in String Theory}

\author{P.~Matlock}
\email{pwm@imsc.res.in}
\affiliation{The Institute of Mathematical Sciences, Chennai, India}
\author{R.~Parthasarathy}
\email{sarathy@imsc.res.in}
\affiliation{The Institute of Mathematical Sciences, Chennai, India}

\begin{abstract}
Buscher duality is a sigma-model duality, implemented by transformation
of the target space. Not only in the case of a flat target space,
but in a general background, should the Buscher duality reduce
to the T-duality familiar in the flat-space string context. 
We exhibit this reduction explicitly using a pp-wave background as 
a tractable example. String theory is solved in a compactified Nappi-Witten 
background and the Buscher-dual theory is likewise solved.
The Hamiltonian is computed in both cases, and the results are 
verified to be T-dual.
\end{abstract}

\maketitle

\section{Overview}
\label{sintro}

It has long been known that string theory compactified on a circle of 
radius $R$ is dual to string theory compactified on a circle of radius $1/R$.
Since this transformation involves a target space compactified on a torus, 
this mapping between two theories is termed T-duality.

When string theory is compactified on $S^1=T^1$, strings not only have resultant 
Kaluza-Klein momentum, but can wind around this circle.
Using the sigma-model approach, it can thus be seen at the level of the 
Hamiltonian that the states in the two theories obtained
by $R$- and $1/R$-compactification may be mapped to one another by 
swapping the r\^ole of the winding and momentum numbers. 
This is T-duality in its simplest incarnation;
the concept may be extended by effecting compactification in more than one 
direction, that is to say, considering string theory on a torus $T^n$.
T-dualities along the toroidal directions form a group of transformations 
of the string theory.

When the target space is flat, there is no difficulty in compactifying the space-time
and using the string theory to understand the duality, since the sigma model in a 
flat background is a free theory. In the case of a curved target space, there is a 
different approach to duality which takes into account the curvature.
Buscher \cite{Buscher,Buscher2} showed how two string theories could be understood to be dual
to each other in a curved background, supplemented with a $B$-field. An explicit 
transformation of the background was exhibited which mapped the sigma model to its 
dual theory. The construction involves identifying a Killing vector, translations 
along which leave the entire action invariant. The Buscher transformation may be
thought of as a certain `metric inversion' along this direction, analogous to the 
inverse radius mentioned above. We refer the reader to Buscher's original
paper for details \cite{Buscher}.

Now, the Buscher duality transformation does not require that the string theory
be compactified, only that the direction chosen be a Killing direction.
Nevertheless, given such a Killing direction along which the action is invariant, 
one may compactify the theory and ask whether a T-duality can be formulated even 
though the target space may have a nontrivial metric and $B$ field.
As hinted in the above paragraph, the answer to this question is that the 
corresponding transformation of the background is the Buscher transformation;
\emph{if two theories are Buscher-dual to each other along a certain Killing direction,
then their compactified counterparts will be T-dual to each other.}
This is shown in the diagram \eqref{diag}. The action ${S}$ is the
sigma-model action in the original background, $S=S_{G,B}[X]$. The Buscher 
dual action is $\busch{S}[X] \equiv S_{\busch{G},\busch{B}}[X]$.
\begin{equation}
\label{diag}
\xymatrix{
 S[X] \ar@{<->}[rr]^{\Pcm{B}} \ar[dd]_{S^1} & & \busch{S}[X] \ar[dd]^{S^1} \\
   & &  \\
 H_c \ar@{<->}[rr]^T & & \busch{H}_c
}
\end{equation}
Along the top of the diagram, the Buscher transformation $\Pcm{B}$
relates the two actions. It can be shown that this transformation is 
idempotent, $\Pcm{B}^2=1$, so there is nothing trivially `beyond the 
diagram' for the Killing direction we have chosen.
Upon compactification on $S^1$,
a Hamiltonian $H_c$ for either theory can be obtained. $H_c$ will have
contributions associated with the energy of winding and momentum modes
in the compact direction, and $H_c$ and $\busch{H}_c$ are expected to
be related in the usual T-dual way; the map from one to the other
involves interchanging the winding and momentum modes so that
$\busch{w}=m$ and $\busch{m}=w$, and inverting the compactification
radius, $\bucsh{R}=\alpha'/2\pi R$.

Thus, the inversion of the radius of the compactified dimension in
flat-space T-duality is a specific instance of the more general
inversion implicit in the Buscher transformation.  For an elementary
introduction to T-duality, the reader may consult \cite{9906108}.
A more complete discussion including the relation with Buscher duality
and many more references may be found in the comprehensive reviews 
\cite{9512181}, \cite{9410237} and \cite{9401139}.

The transformation of Buscher, being in terms of the background metric $G$, 
two-form field $B$ and dilaton $\Phi$ has been investigated extensively in the 
supergravity limit \cite{9807127,9512181,0111072}, and it has been applied 
additionally to questions of time-like T-duality \cite{9807127}, further 
weaving the web of dualities of possible string theories.
Obviously, direct tests of Buscher duality and the relation
to T-duality in the sigma-model context are possible only when both worldsheet 
theories are soluble - not often the case in a nontrivial background - 
and so tests of Buscher duality have largely been limited to supergravity.
Recently, time-like Buscher duality was investigated at the supergravity level in the 
analysis of two pp-wave backgrounds in \cite{0405193}. 
However, as has been extensively discussed in the literature in recent years,
the full string sigma-model may be analysed in such a situation since
a pp-wave represents a special case where the string theory can be
solved in light-cone gauge. It should then be possible to see the T-duality explicitly.
String T-duality in a parallelisable pp-wave background had been considered in \cite{0304169},
and in a maximally-supersymmetric plane-wave background in \cite{0302020}.
However, the close connection between Buscher duality and T-duality was not considered
in this case.

In the present paper we examine this connection between T-duality and Buscher duality, 
making an example of a particular pp-wave background, the 
Nappi-Witten metric \cite{9310112}. This background has been considered in the literature
for both the bosonic \cite{9503222} and NSR string \cite{0305028}.
Possessing Killing directions, it may be compactified, and our first step is
to solve string theory in this compactified background. 
In section \ref{snwspec} we quantise this theory and find the Hamiltonian.
Next, in section \ref{sbuschspec} we find the Buscher-dual background and
worldsheet theory, and quantise it along the same lines.
In section \ref{std} we compare the results and conclude that
T-duality is manifest after the compactification, verifying that
the compactified version of the more general Buscher transformation is
the T-duality transformation.


\section{String spectrum in the compactified Nappi-Witten background}
\label{snwspec}
In this section we begin with the Nappi-Witten background \cite{9310112}
and compactify along a particular Killing direction for the 
metric and also the $B$ field, as explained in the appendix.
We then quantise the string sigma model in light-cone gauge 
and find the spectrum. For simplicity, we devote our attention
to the bosonic sector.

Our notation is as follows.
The background will be ten-dimensional with space-time
indices $\mu,\nu,\dots$..
Of these ten coordinates, we will use $i,j$ indices
to refer to the $1,2$ directions. 
$A,B$ indices will account for six more coordinates,
while the remaining two will be denoted $u$ and $v$.
$\epsilon^{ij}$ is defined by
$\epsilon^{21}=\epsilon_{12}=+1$, and
conventions for worldsheet quantities are
$\eta_{\sigma\sigma}=-\eta_{\tau\tau}=+1$ and
$\epsilon^{\tau\sigma}=\epsilon_{\sigma\tau}=+1$.

\subsection{Equations of motion and mode expansion}
\label{sseom}
The Nappi-Witten background with time-dependent 
$B$-field \cite{9310112} is given by
\begin{eqnarray}
G&=&{\td x^A}^2  + {\td x^i}^2 -2 \td u \td v + (b- \frac{{x^i}^2}{4})\td u^2 ,\\
B&=&\frac12 \epsilon_{ij} x^i \td x^j \wedge \td u
.\end{eqnarray}
We make the choice $b=0$ and transform to new coordinates $y^i$ and $w$ \cite{0203140},
\begin{eqnarray}
x^1 &=& y^1 \cos \frac{u}{2} + y^2 \sin \frac{u}{2} ,\\
x^2 &=& -y^1 \sin \frac{u}{2} + y^2 \cos \frac{u}{2} ,\\
v&=& w-\frac{y^1y^2}{2}
,\end{eqnarray}
so that the background becomes
\begin{eqnarray}
\label{GandB}
G&=&{\td x^A}^2  + {\td y^i}^2 -2 \td u \td w + 2y^2 \td y^1 \td u ,\\
B&=&\frac12 \epsilon_{ij} y^i \td y^j \wedge \td u
.\end{eqnarray}
We note that $B$ is form-invariant under this transformation.
$y^1$ is a Killing direction for $G$, and we now make a gauge transformation 
$B \rightarrow B + \td \lambda$ so that \cite{9503222,0305028}
\begin{equation}
\label{gaugeB}
B= u \td y^1 \wedge \td y^2
.\end{equation}
Now, in accordance with the definition in the appendix, $y^1$ is a Killing 
direction for $B$ also, as well as for $G$.
The bosonic part of the NSR action is given by
\begin{equation}
2\pi\alpha'S=\int\td^2\sigma \left(
  -\frac12 \sqrt{h} h^{\alpha\beta}\partial_\alpha X^\mu \partial_\beta X^\nu G_{\mu\nu} 
  +\frac12 \epsilon^{\alpha\beta}\partial_\alpha X^\mu \partial_\beta X^\nu B_{\mu\nu} 
 \right)
.\end{equation}
Substituting the background $G$ and $B$ into the action, we have
\begin{equation}
\label{firstact}
2\pi\alpha'S=\int\td^2\sigma \left(
  -\frac12 \eta \partial x^A \partial x^A 
  -\frac12 \eta \partial y^i \partial y^i 
  -y^2 \eta \partial y^1 \partial u
  +\eta \partial u \partial w
  +u \epsilon \partial y^1 \partial y^2  
 \right)
.\end{equation}   
This corresponds to the upper-left corner of the diagram \eqref{diag}.
Variation with respect to $w$ gives the equation of motion $\partial^2 u =0$; 
we make the standard light-cone gauge choice $u=u_0 + p^+\tau$ and
obtain the light-cone gauge action
\begin{equation}
\label{lcgA}
2\pi\alpha'S_\textup{LC}=\int\td^2\sigma \left(
  -\frac12 \eta \partial x^A \partial x^A 
  -\frac12 \eta \partial y^i \partial y^i 
  +p^+ y^2 \partial_\tau y^1 
  -p^+  \partial_\tau w
  + u \epsilon^{\alpha\beta} \partial_\alpha y^1 \partial_\beta y^2
 \right)
.\end{equation}   
The term $p^+ \partial_\tau w$ is a total $\tau$ derivative and we discard it.
Now, the remaining equations of motion are as follow:
\begin{eqnarray}
\delta x^A &\rightarrow& \partial^2 x^A =0 ,\\
\delta y^1 &\rightarrow& \partial^2 y^1 = p^+(\partial_\tau+\partial_\sigma) y^2 ,\\
\delta y^2 &\rightarrow& \partial^2 y^2 = -p^+(\partial_\tau+\partial_\sigma) y^1
.\end{eqnarray}
The first leads to the standard mode expansions for $x^A$ and need not be discussed.
The next two are decoupled by combination into a single complex 
equation for the quantity $Y(\sigma,\tau) \equiv y^1 + i y^2$.
\begin{equation}
\label{Yeom}
\big(\partial^2 + i p^+(\partial_\tau + \partial_\sigma)\big)Y(\sigma,\tau)=0
\end{equation}
A general form for $Y=Y_0(\sigma,\tau)$ with homogeneous boundary condition $Y_0(2\pi,\tau)=Y_0(0)$ is
\begin{equation}
Y_0(\sigma,\tau)=\sum_{n\in{\mathbb Z}} y_n(\tau) e^{in\sigma}
.\end{equation}
Substituting into the equation of motion, we find 
$y_n(\tau) = a_n e^{-in\tau} + \tilde{a}_n e^{i(p^+ + n)\tau}$; 
$a_n$ and $\tilde{a}_n$ will be left- and right-movers.
To this solution we wish to add a particular solution to accommodate 
compactification and thus closed-string winding modes along the $y^1$-direction,
so that the boundary condition
\begin{equation}
\label{YBC}
Y(2\pi,\tau)-Y(0,\tau)=2\pi Rw 
,\qquad \forall \tau
\end{equation}
is appropriate.
Here, $R$ is the compactification radius, and $w \in {\mathbb Z}$ is the winding number.
The complete expansion for $Y$ is then
\begin{equation}
\label{Ymodeexp1}
Y(\sigma,\tau)= Rw(\sigma-\tau) 
                + \sum_{n\in{\mathbb Z}} a_n e^{in(\sigma-\tau)}
                + e^{ip^+\tau} \sum_{n\in{\mathbb Z}} \tilde{a}_n e^{in(\sigma+\tau)}
.\end{equation}

\subsection{Space-time momentum and zero modes}
\label{sskkmzm}
Compactification of $y^1$ implies quantisation of the space-time momentum 
conjugate to $y^1$, and this must be found in terms of worldsheet degrees 
of freedom. 
Space-time transformation symmetries are global worldsheet symmetries;
we use the Noether procedure (as in \cite{GSW}, pg.69) to find the 
conserved current associated with $y^1$ translation. We obtain
\begin{equation}
2\pi\alpha' J^\alpha = -\eta^{\alpha\beta} \partial_\beta y^1
   +\eta^{\alpha\beta} y^2 \partial_\beta u
   +u \epsilon^{\alpha\beta} \partial_\beta y^2
.\end{equation}
The space-time momemtum $P^1$ corresponding to $y^1$-translation is thus
\begin{equation}
\label{KKmom}
P^1 \equiv \int \td \sigma J^\tau 
 =\frac1{2\pi\alpha'} \int_0^{2\pi} \td \sigma \big( 
                \partial_\tau y^1 + p^+ y^2 + u \partial_\sigma y^2
                                      \big)
.\end{equation}
Substituting the mode expansion for $Y=y^1+iy^2$, we obtain
\begin{equation}
\label{KKmomz}
P^1=\frac{1}{\alpha'}\left[ -\frac{i}{2}p^+(a_0-a_0^\dagger) - wR \right]
\end{equation}
which is real (Hermitian) and independent of $\tau$, as we expect.
Inserting this in our mode expansion of \er{Ymodeexp1} we arrive at
\begin{equation}
\label{Ymodeexpz}
Y(\sigma,\tau)=Rw\sigma + \big( 
               \alpha' P^1 +\frac{i}{2}p^+ (a_0 - a_0^\dagger)
                          \big)\tau
             + \sum_{n\in\mathbb{Z}}
                \left( 
                  a_n e^{in(\sigma-\tau)}
                  + e^{ip^+\tau} \tilde{a}_n e^{in(\sigma+\tau)} 
                \right)
.\end{equation}
Before continuing on to quantisation, we remark that if we were to choose
a `standard' Kaluza-Klein condition $P^1=m/2\pi R$, we would not obtain conventional 
T-duality from this mode expansion, due to the extra term linear in $\tau$.
We could, of course, have taken a different route and substituted for 
the $\sigma$ coefficient instead of the $\tau$ coefficient, but the 
boundary condition \eqref{YBC} is unambiguous and leads us to the same 
result in any case.
We will elaborate on this at the following section and in section \ref{std}.

\subsection{Quantisation and spectrum}
\label{ssqas}
The canonical momenta conjugate to the $y^i$ coordinates may be read off from \er{firstact},
\begin{equation}
\label{canmom}
\pi_1=\partial_\tau y^1 + u \partial_\sigma y^2 + y^2 \partial_\tau u
,\qquad
\pi_2=\partial_\tau y^2 - u \partial_\sigma y^1 
.\end{equation}
Quantisation corresponds to imposing the commutation relation
\begin{equation}
[ y^i(\sigma,\tau) , \pi_j(\sigma',\tau)  ] = i \pi \delta^i_j \delta(\sigma-\sigma')
.\end{equation}
As mentioned when we obtained the equations of motion in section \ref{sseom},
we will ignore the `flat' directions $x^A$.
For the oscillator modes, the above commutation relation leads to\footnote{The 
reader is reminded that, in contrast to the usual flat-space notation,
$a_n$ and $a_{-n}$ are distinct operators unrelated by Hermitian conjugation.}
\begin{equation}
[a_n , a_n^\dagger ] = [\tilde{a}_{n}^\dagger , \tilde{a}_{n} ] = \frac{1}{p^+}
,\qquad \forall n\ne0 
.\end{equation}
For the `zero modes' (non-oscillatory), we have
\begin{equation}
[P^1,a_0]=\frac{i}{2\alpha'}
, \qquad [\tilde{a}_0^\dagger,\tilde{a}_0]=\frac{1}{p^+}   
.\end{equation}
All other commutators (except of course those obtained by Hermitian
conjugation from the above) vanish. 
As a consequence of these `zero-mode' commutators, we have that
\begin{equation}
\label{ncyy}
[ y^1,y^2] = \frac{1}{2ip^+} , \qquad \textup{and} \qquad [\pi_1,\pi_2]=0
.\end{equation}
This non-commutativity of the coordinates has been seen in similar
situations \cite{0302020,0304169}.

The Hamiltonian may be written using the Lagrangian \eqref{lcgA}
and the momenta \eqref{canmom}. The result is
\begin{equation}
\label{pHamil}
2\pi\alpha' H = \frac12 \int_0^{2\pi} \td \sigma \big( 
                 \partial_\tau Y\partial_\tau Y^\dagger
                 +\partial_\sigma Y\partial_\sigma Y^\dagger
                 \big)
.\end{equation}
Substituting the mode expansion for $Y$ and performing the integration
we arrive at
\begin{eqnarray}
\label{Hamil}
\alpha' H &=&
      (Rw)^2 
      + \big( \alpha' P^1 + \frac{i}2 p^+ (a_0-a_0^\dagger) \big)^2
    \nn\\
    &&{} +\frac12 p^+ \tilde{a}_0\tilde{a}_0^\dagger 
     + \sum_{n>0} n^2 \big( 
        a_na_n^\dagger + \tilde{a}_n\tilde{a}_n^\dagger
      + a_{-n}a_{-n}^\dagger + \tilde{a}_{-n}\tilde{a}_{-n}^\dagger
                               \big)
,\end{eqnarray}
which is lower-left corner of \eqref{diag}.

We notice at this point that in order for this theory to exhibit self T-duality,
we seem to be forced to make one of two choices. 
One choice would be to first restrict $a_0-a_0^\dagger$, the $y^2$-coordinate zero mode,
to vanish. The space-time momentum $P^1$ may then be quantised in the usual Kaluza-Klein way,
so that $P^1=m/2\pi R$, and the standard T-duality involving an exchange of momentum 
number $m$ and winding number $w$ coupled with an inversion of the compactification radius
would obtain.
As we see no compelling reason why the theory should apply only 
to strings at $y^2=0$, we instead make a different choice.
We will interpret the quantity $ \alpha' P^1 + \frac{i}2 p^+ (a_0-a_0^\dagger) $ as
the `Kaluza-Klein' momentum, and thus set
\begin{equation}
\label{kkm}
 \alpha' P^1 + \frac{i}2 p^+ (a_0-a_0^\dagger) = \frac{m}{2\pi R}
.\end{equation}
Further comments will be made in section \ref{std}.
Of course, in the present paper we are not interested in T-duality
within this theory alone, but between this and its Buscher dual,
to which we now turn our attention.

\section{String spectrum in the Buscher-dual background}
\label{sbuschspec}
Here, we first find the Buscher dual to the background
considered in section \ref{snwspec}. We quantise the resulting
sigma model along the same lines so that the mode expansion
and spectrum may be compared. 

\subsection{The Buscher-dual background}
\label{ssbuscher}
To perform a Buscher transformation, we must have a Killing vector
common to the $G$ and $B$ fields, as defined in the appendix, and
this is also a requirement for a consistent compactification.
Here we assume that the coordinates are chosen so that one of them defines
this Killing direction. We denote this `Killing coordinate' by the index $0$
and all other coordinates with the indices $\theta$ and $\phi$.
The Buscher transformation is derived by gauging the translation
symmetry of the Killing coordinate.
From \cite{Buscher}, the Buscher transformation of the background 
$G$ and $B$ into $\busch{G}$ and $\busch{B}$ is then given by
\begin{eqnarray}
\label{buschtran}
\busch{G}_{00} = \frac{1}{G_{00}} ,&\qquad&
\busch{G}_{\theta\phi} =
   G_{\theta\phi} - \frac{G_{0\theta}G_{0\phi} - B_{0\theta}B_{0_\phi} }{G_{00}} ,\nn\\
\busch{G}_{0\theta} = \busch{G}_{\theta 0} = \frac{B_{0\theta}}{G_{00}} ,&\qquad&
\busch{B}_{\theta\phi} = 
   B_{\theta\phi} + \frac{G_{0\theta}B_{0\phi} - B_{0\theta}G_{0_\phi} }{G_{00}} ,\nn\\
\busch{B}_{0\theta} = - \busch{B}_{\theta 0} = \frac{G_{0\theta}}{G_{00}} ,&\qquad&
\busch{\Phi} = \Phi - \frac12 \log |G_{00}|
.\end{eqnarray}
We have included the dilaton transformation $\Phi\rightarrow\tilde{\Phi}$
for completeness. Although we will not need it for our purpose of finding the spectrum,
we will see that it becomes important when comparing the spectra of the two theories
since in principle $e^{-2\Phi}$ multiplies the whole action.

We now begin with our sigma-model background of equations \eqref{GandB} and \eqref{gaugeB}
which we used in section \ref{snwspec}, and with $y^1$ the Killing coordinate
we obtain the dual background
\begin{eqnarray}
\label{dualBG}
\busch{G} &=& {\td x^A}^2 + (\td y^1)^2 + 2 u \td y^1 \td y^2 
               + (1+u^2) (\td y^2)^2 -2 \td u \td w 
              -(y^2)^2 {\td u}^2 ,\\
\busch{B} &=& y^2 \td y^1 \wedge \td u - u y^2 \td y^2 \wedge \td u
.\end{eqnarray}
The action is then 
\begin{eqnarray}
\label{buschact}
2\pi\alpha'\busch{S}&=&\int\td^2\sigma \bigg(
  -\frac12 \eta \partial x^A \partial x^A 
  -\frac12 \eta \partial y^1 \partial y^1 
  - \frac12 (1+u^2) \eta \partial y^2 \partial y^2
  +\eta \partial u \partial w \nn\\
 &&\qquad\qquad{}
  + u \eta \partial y^1 \partial y^2
  + \frac12 y^2 \eta \partial u\partial u
  + y^2 \epsilon \partial y^1 \partial u  
  + u y^2 \epsilon \partial u \partial y^2  
 \bigg)
,\end{eqnarray}   
which represents the upper-right corner of \eqref{diag}.
Once again, the gauge $u=u_0 + p^+ \tau$ may be chosen, 
and the resultant light-cone gauge action is given by
\begin{eqnarray}
\label{BuschAct}
2\pi\alpha' \busch{S}_\textup{LC} &=& \int \td^2\sigma \bigg(
   {}-\frac12 \eta \partial x^A\partial x^A 
   -\frac12 \eta\partial y^1\partial y^1 - u \eta \partial y^1\partial y^2
   -\frac12 (1+u^2) \eta \partial y^2 \partial y^2
   \nn\\
   && \qquad\qquad{}
   -p^+\partial_\tau w -\frac12 {p^+}^2(y^2)^2 
   -p^+y^2\partial_\sigma y^1 + u p^+ y^2 \partial_\sigma y^2
\bigg)
.\end{eqnarray}

\subsection{Equations of motion and mode expansion}
\label{ssdeom}
Varying the action of \er{BuschAct} gives equations of motion for $y^i$ which may be written
\begin{eqnarray}
\partial^2 (y^1 + u y^2) &=& - p^+ (\partial_\tau + \partial_\sigma) y^2 \\
-\partial^2 y^2 &=& p^+ (\partial_\tau + \partial_\sigma) (y^1 + u y^2)
.\end{eqnarray}
Making the definition
$\busch{Y}=y^1+uy^2 + i y^2$
we find that the equation of motion for $\busch{Y}$ coincides with
the equation of motion \eqref{Yeom} for the analogous $Y$ of 
section \ref{sseom}, with the replacement $p^+ \rightarrow -p^+$.
Again we wish to impose a periodic boundary condition with winding number
on $y^1$ alone, and \er{YBC} remains correct here as well. 
We thus can carry over the mode expansion of \er{Ymodeexp1} and we have
\begin{equation}
\label{Ymodeexp2}
\busch{Y}(\sigma,\tau)= \bucsh{R}\busch{w}(\sigma-\tau)
                + \sum_{n\in{\mathbb Z}} b_n e^{in(\sigma-\tau)}
                + e^{-ip^+\tau} \sum_{n\in{\mathbb Z}} \tilde{b}_n e^{in(\sigma+\tau)}
,\end{equation}
where $\busch{w}$ is the winding number and $\bucsh{R}$ is the radius 
of compactification of the $y^1$ coordinate, but of course measured in 
the new Buscher-dual background.

\subsection{Space-time momentum and zero modes}
\label{ssdkkmzm}
Following the same procedure as in section \ref{sskkmzm} we find the conserved current
associated with $y^1$-translations to be
\begin{equation}
\label{dKKcurr}
2\pi\alpha' \busch{J}^\alpha = -\eta^{\alpha\beta}(\partial_\beta y^1 + u \partial_\beta y^2)
                       -p^+ y^2 \eta^{\alpha\sigma}
\end{equation}
so the Kaluza-Klein momentum is
\begin{equation}
\label{KKdmom}
\busch{P}^1=\frac1{2\pi\alpha'} \int_0^{2\pi} \td \sigma 
                \big( 
                \partial_\tau (y^1 + u y^2) - p^+ y^2 
                \big)
           =\frac{1}{\alpha'}\left[ \frac{i}{2}p^+(b_0-b_0^\dagger) 
             - \busch{w}\bucsh{R} \right]
,\end{equation}
which is clearly analogous to \er{KKmomz}.
We substitute into the mode expansion \eqref{Ymodeexp2} to arrive at
\begin{equation}
\label{Ydmodeexpz}
\busch{Y}(\sigma,\tau)=\bucsh{R}\busch{w}\sigma 
       + \big( \alpha'\busch{P}^1 
       - \frac{i}{2}p^+ (b_0 - b_0^\dagger) \big) \tau 
                + \sum_{n\in\mathbb{Z}}
                \left( 
                  b_n e^{in(\sigma-\tau)}
                  + e^{-ip^+\tau} \tilde{b}_n e^{in(\sigma+\tau)} 
                \right)
.\end{equation}

\subsection{Quantisation and spectrum}
\label{ssdqas}
The $y^i$ coordinates are not the correct canonical variables for quantisation.
Inspection of the definition of $\busch{Y}$ in section \ref{ssdeom} 
leads us to change variables from $y^i$ to
\begin{equation}
z \equiv \frac{\busch{Y} + \busch{Y}^\dagger}{2} = y^1 + u y^2 
\qquad \textup{and} \qquad 
y \equiv \frac{\busch{Y} - \busch{Y}^\dagger}{2} = y^2
,\end{equation}
in terms of which the action \eqref{BuschAct} becomes
\begin{equation}
2\pi\alpha'\busch{S} = -\frac12 \int \td^2\sigma\bigg(
\eta\partial x^A\partial x^A + \eta \partial z \partial z
+2 p^+ y (\partial_\tau + \partial_\sigma) z 
+\eta \partial y\partial y 
-4 p^+ u y \partial_\sigma y
\bigg)
.\end{equation}
We neglect the the last term, a total $\sigma$ derivative.
The canonical momenta are now given by
\begin{equation}
\pi_z = \partial_\tau z - p^+ y
\qquad \textup{and} \qquad 
\pi_y = \partial_\tau y
.\end{equation}
We quantise by requiring
\begin{equation}
[ z(\sigma,\tau) , \pi_z(\sigma',\tau)  ] =
[ y(\sigma,\tau) , \pi_y(\sigma',\tau)  ] = i \pi \delta(\sigma-\sigma')
,\end{equation}
which implies that modes of $\busch{Y}$ in \er{Ydmodeexpz} have non-vanishing commutators
\begin{eqnarray}
[b_n,b_n^\dagger]=[\tilde{b}_n^\dagger,\tilde{b}_n]=-\frac{1}{p^+} 
\quad \forall n \ne 0 
\qquad \textup{and} \qquad
[\busch{P^1},b_0]=\frac{i}{2\alpha'}
, \qquad [\tilde{b}_0^\dagger,\tilde{b}_0]=-\frac{1}{p^+}
.\end{eqnarray}
Similar to \er{ncyy}, it can be seen that the coordinates $z$ and $y$ do not commute, 
which implies that $[y^1,y^2]\ne0$ in this case as well.

The Hamiltonian is given by
\begin{equation}
\label{Hamild}
2\pi\alpha' H = \frac12 \int_0^{2\pi} \td \sigma \big( 
                 \partial_\tau \busch{Y}\partial_\tau \busch{Y}^\dagger
                 +\partial_\sigma \busch{Y}\partial_\sigma \busch{Y}^\dagger
                 + p^+ y \partial_\sigma z
                 \big)
.\end{equation}
Apart from the last term, this Hamiltonian is exactly analogous to \er{pHamil}.
The last term is in fact seen not to contribute when 
written using the mode expansion for $\busch{Y}$; the result is therefore
identical in form to the Hamiltonian in the dual case, \er{Hamil},
\begin{eqnarray}
\label{bHamil}
\alpha' \busch{H} &=&
      (\bucsh{R}\busch{w})^2 
      + \big(  \alpha'\bucsh{P}^1   - \frac{i}2 p^+ (b_0 - b^\dagger_0) \big) ^2
    \nn\\
    &&{} -\frac12 p^+ \tilde{b}_0\tilde{b}_0^\dagger 
     + \sum_{n>0} n^2 \big( 
        b_n b_n^\dagger + \tilde{b}_n\tilde{b}_n^\dagger
      + b_{-n}b_{-n}^\dagger + \tilde{b}_{-n}\tilde{b}_{-n}^\dagger
                               \big)
.\end{eqnarray}
Thus we have the final, lower-right corner of \eqref{diag}.
In the Buscher-dual case, we are compelled to make the same interpretation as we
did leading up to equation \eqref{kkm}. That is, we identify
the coefficient of $\tau$ as the `Kaluza-Klein' momentum and give it a momentum number,
\begin{equation}
\label{bkkm}
\alpha'\bucsh{P}^1   - \frac{i}2 p^+ (b_0 - b^\dagger_0) =
 \frac{\busch{m}}{2\pi \busch{R}}
.\end{equation}


\section{T-duality}
\label{std}

We have defined discrete momentum numbers in equations \eqref{kkm} and \eqref{bkkm}.
Although the momentum numbers $m$ and $\busch{m}$ are not conventional Kaluza-Klein
momenta, in the sense that the quantities $P^1$ and $\busch{P}^1$ themselves are
the conserved space-time momenta in the compact direction, we see from our expressions
for the Hamiltonian in either case that if self T-duality is to hold, then the
momentum number so defined is the T-dual of the winding number.
The centre-of-mass mode of the $y^2$ coordinate is thus mixed into the definition
of the momentum number of the $y^1$ coordinate. This is a 
geometrical consequence of the term $y^2 \td y^1 (\wedge) \td u$ in the metric \eqref{GandB}
and the $B$-field in \eqref{dualBG}; the conserved space-time momenta of \er{KKmomz} and \er{KKdmom}
along the Killing direction depend on the centre-of-mass position in the $y^2$ direction.
This mixing term in the background can also be held responsible for 
the non-commutativity that we found between $y^1$ and $y^2$ as in \er{ncyy}.

Our task is to understand the bottom edge of our diagram \eqref{diag} and to exhibit the 
expected T-duality between the compactified theories.
Looking at the two Hamiltonians \eqref{Hamil} and \eqref{bHamil} corresponding 
to the bottom corners of the diagram, we see that they na\"ively do 
not appear to be T-dual to each other; in fact they appear to be very similar.

Let us recall, however, that the two Hamiltonians are expressed in different backgrounds.
Following \cite{9401139}, the T-duality may be clearly seen as follows.
$2\pi R$ is simply a coordinate distance along the compactified $y^1$-direction; from 
the Buscher transformation, distance in these two metrics is inversely related, as in 
the expression for $\busch{G}_{00}$ in \er{buschtran}.
Considering this direction, let us define the $y^1$-coordinate as dimensionless, 
by scaling $y^1 \rightarrow R y^1$, and the metric will now have dimensions
of length-squared, $G_{00}$ is scaled by $R^2$. 
We will redefine $R$ to be the dimensionless radius, so $R\rightarrow R/\sqrt{\alpha'}$.
In order to formulate the Buscher transformation in terms of a dimensionless
metric, we instead let $G$ remain dimensionless by scaling with $\alpha'$.
In summary, we have the redefinition
\begin{equation}
y^1 \rightarrow R y^1 ,\qquad \textup{and} \qquad G_{00} \rightarrow \frac{R^2}{\alpha'}G_{00}
.\end{equation}
Then, as shown in \cite{9401139}, the Buscher transformation implies
\begin{equation}
G_{00} \rightarrow \busch{G}_{00}=\frac{1}{G_{00}}, \qquad \textup{and} \qquad \busch{R}=\frac1{R}
.\end{equation}
We see it is also necessary to change the sign of $p^+$, but this was 
just a gauge choice for the $u$ coordinate;
it may be chosen with the opposite sign in the dual case.

The non-zero modes of the two Hamiltonians should coincide.
However, the Hamiltonian has dimensions of energy, and should 
be measured in the correct metric as well, against $G_{00}$. 
This metric is different for the two Hamiltonians. 
The resolution of this problem (which is not obvious in our 
expressions since in our case $G_{00}=1$) is through the dilaton.
The dilaton factor of $e^{-2\Phi}$ which would be in the action if
we had considered the dilaton field shifts under the Buscher 
transformation \eqref{buschtran}, producing a factor of $G_{00}$
and rendering the modes (this argument applies to the zero-modes as well) 
comparable in the two systems;
they may be taken `at face value' in the expressions \eqref{Hamil} 
and \eqref{bHamil}.

We have thus shown an explicit example of T-duality in a curved 
background metric with non-trivial $B$-field, and demonstrated
how it arises in the context of Buscher duality when compactified
along the Killing direction.


\section*{Appendix - generalised Killing equation}

We make a simple but convenient extension to the definition of a Killing vector.
For an arbitrary tensor $\lambda_{\mu\nu}$, let us consider an 
infinitesimal coordinate shift $x^\mu \rightarrow {x'}^\mu = x^\mu + \xi^\mu(x)$.
The requirement that $\lambda_{\mu\nu}$ be form invariant under this 
transformation, so that $\lambda'_{\mu\nu}$ is the same function 
of $x'$ as $\lambda_{\mu\nu}$ is of $x$, leads to the equation
\begin{equation}
\label{kill}
\xi^\sigma \nabla_\sigma\lambda_{\mu\nu} 
+ \lambda_{\mu\sigma}\nabla_\nu \xi^\sigma 
+ \lambda_{\sigma\nu}\nabla_\nu \xi^\sigma =0
.\end{equation}
This `generalised Killing equation' may be extended to an arbitrary-rank 
tensor in the obvious way.
Equation \eqref{kill} of course gives the usual Killing equation 
for $\xi$ when $\lambda$ is taken to be the metric. 
Using this equation, we may talk of a `Killing vector for $B$'
and by this mean a solution to \er{kill} with $\lambda=B$.
As usual, if the coordinates are chosen so that the $0$-direction
coincides with the Killing direction, then $\xi^\mu=\delta_0^\mu$
and \er{kill} reduces to $\partial_0 \lambda_{\mu\nu}=0$.
In the present paper, this corresponds to our condition
that the metric and $B$-field have no $y^1$-dependence,
which we require for both the compactification and the Buscher 
transformation.


\end{document}